\documentclass[12pt]{iopart}

\usepackage{graphicx}
\usepackage{here}

\usepackage{stackrel}
\usepackage{amssymb,accents,latexsym}
\usepackage{braket}

\usepackage{subfigure}
\usepackage{tikz}

\newcommand{\TF}{{\textrm{\tiny TF}}}

\begin{document}

\title[Tsallis statistics and thermofractals: applications to high energy and hadron physics]{Tsallis statistics and thermofractals: applications to high energy and hadron physics}

\author{Eugenio~Meg\'{\i}as$^{1}$, Evandro Andrade II$^{2}$, Airton Deppman$^{3}$, Arnaldo Gammal$^{3}$, D\'ebora P. Menezes$^{4}$, Tiago Nunes da Silva$^{4}$ and Varese S. Tim\'oteo$^{5}$}

\address{
$^{1}$ Departamento de F\'{\i}sica At\'omica, Molecular y Nuclear and Instituto Carlos I de F\'{\i}sica Te\'orica y Computacional, Universidad de Granada, \\ 
Avenida de Fuente Nueva s/n, 18071 Granada, Spain \\
$^{2}$ Departamento de Ci\^encias Exatas e Tecnol\'ogicas, Universidade Estadual de Santa Cruz, Ilh\'eus, CEP 45662-900 Bahia, Brazil \\
$^{3}$ Instituto de F\'{\i}sica, Universidade de S\~ao Paulo, Rua do Mat\~ao 1371-Butant\~a, S\~ao Paulo-SP, CEP 05580-090, Brazil \\
$^{4}$ Departamento de F\'{\i}sica, CFM-Universidade Federal de Santa Catarina, Florian\'opolis, SC-CP. 476-CEP 88.040-900, Brazil \\
$^{5}$ Grupo de \'Optica e Modelagem Num\'erica, Faculdade de Tecnologia GOMNI/FT - Universidade Estadual de Campinas - UNICAMP 13484-332, Limeira, SP, Brazil 
}
\ead{eoasegundo@uesc.br, deppman@usp.br, gammal@if.usp.br, emegias@ugr.es, debora.p.m@ufsc.br, t.j.nunes@ufsc.br, varese@unicamp.br}
\vspace{10pt}

\begin{abstract}
We study the applications of non-extensive Tsallis statistics to high
energy and hadron physics. These applications include studies of $pp$
collisions, equation of state of QCD, as well as Bose-Einstein
condensation. We also analyze the connections of Tsallis statistics
with thermofractals, and address some of the conceptual aspects of the
fractal approach, which are expressed in terms of the renormalization
group equation and the self-energy corrections to the parton mass. We
associate these well-known concepts with the origins of the fractal
structure in the quantum field theory.
\end{abstract}

%
\vspace{2pc}

\section{Introduction}
\label{sec:intro}

Important advances in the study of the phenomenology of Quantum Chromodynamics (QCD) in the hot and dense regimes, in particular in the quark-gluon plasma, have been developed in recent years. These studies have motivated the introduction of several approaches, including lattices studies~\cite{Borsanyi:2010cj}, chiral quark models~\cite{Fukushima:2003fw,Megias:2004hj}, hadron resonance gas (HRG) models~\cite{Hagedorn:1965st,Hagedorn:1984hz,Huovinen:2009yb,Megias:2012kb}, and holographic models~\cite{Sakai:2004cn,Erlich:2005qh}, among others. Motivated by the large amount of information that emerged from high energy physics (HEP) and heavy-ion physics experiments, the consequences of those advances are far-reaching. Let us summarize the three fundamental theories that will be used for the developments that will be discussed below: the Yang-Mills field (YMF) theory, the fractal geometry, and the non-extensive statistics proposed by Constantino Tsallis.

YMF theory is a prototype theory that allows describing most of the physical phenomena~\cite{Yang:1954ek}. It was incorporated in the electro-weak theory in the 1960s, and in QCD in the 1970s.  One of the fundamental properties of physics laws is the renormalization group (RG) invariance, an aspect that plays an important role in the renormalization properties of YMF theories after the ultraviolet divergences are subtracted~\cite{Dyson:1949ha,GellMann:1954fq}.

Fractals are complex systems presenting a fine structure with an undetermined number of components that are also fractals similar to the original system but at a different scale~\cite{Mandelbrot}. This property is known as self-similarity. Fractal geometry has been used to describe many natural shapes that can be observed in everyday life. A direct consequence of the self-similarity is the power-law behavior of distributions observed for fractals.  

Tsallis statistics was introduced as a generalization of Boltzmann-Gibbs (BG) statistics by considering a non-additive form of the entropy~\cite{Tsallis:1987eu}. Contrary to the exponential distribution of BG statistics, Tsallis distribution has a power-law behavior which has led in the last few years to a wide range of applications apart from HEP, see e.g.~\cite{Tempesta:2011vc,Tsallis:book}. However, the full understanding of this statistics has not been accomplished yet, one of the open questions being the physical origin of the power-law behavior in physical systems.

The goal of this manuscript is to provide an overview of the main applications of Tsallis statistics to HEP and hadron physics, as well as to study the link between RG invariance of YMF theory, fractals and Tsallis statistics. The manuscript is organized as follows. In Sec.~\ref{sec:thermofractals} we will introduce Tsallis statistics and explore the main properties of fractals. We will also investigate the connection between thermofractals and Tsallis statistics, and address the thermofractal description of YMF theory. We will study in Sec.~\ref{sec:applications} some of the recent applications of Tsallis statistics to HEP, including $pp$ collisions, QCD thermodynamics, and Bose-Einstein condensation (BEC). Finally, in Sec.~\ref{sec:conclusions} we present our conclusions.

\section{Tsallis statistics, thermofractals and Yang-Mills fields}
\label{sec:thermofractals}

In this section, we will provide an introduction to Tsallis statistics and the formalism of thermofractals, and explore the link between these two descriptions.

\subsection{Tsallis statistics}
\label{subsec:Tsallis_statistics}

Tsallis statistics is a generalization of BG statistics, with entropy given by~\cite{Tsallis:1987eu}
\begin{equation}
S_q \equiv - k_B \sum_i p(x_i)^q \ln_q^{(-)} p(x_i)  \,, \label{eq:Sq}
\end{equation}
where $p(x)$ is the probability of $x$ to be observed, $k_B$ is the Boltzmann constant, and $q$ is the entropic index that quantifies how Tsallis entropy departs from the extensive BG statistics. Tsallis statistics is defined in terms of the $q$-exponential and $q$-logarithmic functions, given by
\begin{equation}
e_q^{(\pm)}(x)=[1 \pm (q-1)x]^{\pm \frac{1}{q-1}}\,, \qquad \ln^{(\pm)}_q(x)=\pm \frac{x^{\pm(q-1)}-1}{q-1} \,,
\end{equation}
respectively. A consequence of Eq.~(\ref{eq:Sq}) is that the entropy of the system is non-additive, i.e. for two independent systems $A$ and $B$~\cite{Tsallis:1987eu}
\begin{equation}
S_{A+B} = S_A + S_B + k_B^{-1} (q-1)S_A S_B \,.
\end{equation}
Notice that $e_q^{(\pm)}(x) \stackrel[q \to 1]{\longrightarrow}{} e^x$ and $\ln_q^{(\pm)}(x) \stackrel[q \to 1]{\longrightarrow}{} \ln(x)$, so that as $q \rightarrow 1$ the BG statistics is recovered and the entropic form becomes additive.

\subsection{Fractals and self-similarity}
\label{subsec:fractals}

We will  introduce now the main concepts related to fractals. Fractals are defined by their self-similar properties at different scales. A scaling transformation changes the size of a system by a scaling factor, $\lambda$. On the other hand, a self-similar system is a system which is similar to a part of itself. A typical example of a fractal is the  Sierpi\'nski triangle. If length is reduced by the scaling factor as $\ell(\lambda) = \ell_0/\lambda$, then a system of dimension~$D$ can be filled by $N(\lambda) = N_0 \lambda^D$ smaller self-similar systems. Then, one can define the fractal dimension as
\begin{equation}
D \equiv \lim_{\lambda \to \infty} \frac{\ln N(\lambda)}{\ln \lambda} \,.
\end{equation}
This definition is valid for both fractal and non-fractal systems. Another example of a fractal is the length of coastlines, as it depends on the resolution. If $L_0 = N_0 \ell_0$ is the measured length at some initial resolution, $\ell_0$, then the measured length at a better resolution $\ell(\lambda) = \ell_0/\lambda$ turns out to be 
\begin{equation}
L(\lambda) = N(\lambda) \ell(\lambda) = L_0 \cdot \lambda^{D-1} \,.
\end{equation}
One can see that an increase of $L$ with $\lambda$ is indicative of a fractal dimension $D > 1$. For the coastline of Great Britain, it is $D \simeq 1.25$, but generically different shapes induce a fractal spectrum of dimensions.

\subsection{Thermofractals}
\label{subsec:thermofractals}

The emergence of the non-extensive behavior in physical systems has
been attributed in the literature to different causes: i)~long-range
interactions and correlations~\cite{Borland:1998}; ii)~temperature
fluctuations; and iii)~finite size of the system. We will explore
below a natural derivation of non-extensive statistics in terms of
thermofractals. These are systems in thermodynamical equilibrium
presenting the following properties~\cite{Deppman:2016fxs}:
\begin{itemize}
\item The total energy of the system is given by~$U = F + E$, where $F$ is the kinetic energy, and $E$ is the internal energy of $N$ constituent subsystems, so that $E = \sum_{i=1}^N \varepsilon_i^{(1)}$.
\item The constituent subsystems are thermofractals, which means that the energy distribution $P_{\TF}(E)$ is self-similar or self-affine, i.e. at level $n$ of the hierarchy of subsystems, $P_{\TF (n)}(\varepsilon)$ is equal to the distribution in any other level~$P_{\TF (n)}(\varepsilon) \propto P_{\TF (n+ n^\prime)}(\varepsilon)$.
\item At any level $n$ of the fractal structure, the phase space is so narrow that one can consider~$P_{\TF}(E_n) dE_n = \rho \, dE_n$. This means that the internal energy fluctuations are small enough to be disregarded, and then the internal energy can be considered to be equal to the component mass~$m$.
\end{itemize}
Using these properties, it is possible to show that thermofractals result in energy distributions of the kind~\cite{Deppman:2016fxs,Deppman:2019yno,Deppman:2020gbu}
\begin{equation}
P_{\TF(n)}^{(\pm)}(\varepsilon) =  A_{(n)} \, \cdot  e_q^{(\pm)}\left( -\frac{\varepsilon}{k_B \tau} \right) \,, \label{eq:PTFpm}
\end{equation}
so that the energy distribution of thermofractals obeys Tsallis statistics. The positive(negative) version of the q-exponential function corresponds to type-I (type-II) thermofractals. The main difference between the two kinds of thermofractals is the character of the distribution: type-I requires a cut-off because of the negative sign in the argument, while type-II presents a distribution without a cut-off.

\subsection{Fractal structures in Yang-Mills fields}
\label{subsec:Yang-Mills}

We have seen in Sec.~\ref{subsec:thermofractals} that thermofractals obey Tsallis statistics. On the other hand, as we will see in Sec.~\ref{sec:applications}, the phenomenology of QCD can be successfully described by this statistics. Then, a natural question arises: Is it possible a thermofractal description of YMF theories? We will address below this question.

Partons are considered fundamental particles without internal structure, therefore, in principle, they cannot be fractals. This statement holds until the QCD vacuum is not considered, though. We know that vacuum polarization is an essential part of the interaction
not only in QCD but also in Quantum Electrodynamics and YMF theories in general~\cite{Casher:1974vf}. The vacuum structure is an important component of partons
interactions and the parton self-energy~\cite{Casher:1973uf}. We will describe below how the fractal structure appears in parton dynamics.

The YMF theory was shown to be renormalizable in
Ref.~\cite{tHooft:1972tcz}, which means that the regularized vertex
functions are related to the renormalized ones, to which the
renormalized parameters, $\bar{m}$ and $\bar{g}$, are associated
by~\cite{Dyson:1949ha,GellMann:1954fq}
\begin{equation}
\Gamma(p,m,g)=\lambda^{-D} \Gamma(p,\bar{m},\bar{g}) \,,
\end{equation}
where $\lambda$ is the scale transformation parameter, i.e. $p^\mu \to p^{\prime \mu} = \lambda p^\mu$. This property is described by the RG equation, also known as Callan-Symanzik (CS) equation~\cite{Callan:1970yg,Symanzik:1970rt}
\begin{equation}
\left[M\frac{\partial}{\partial M}  + \beta_{\bar g} \frac{\partial}{\partial \bar{g}}  + \bar\gamma \right]\Gamma = 0 \,,
\end{equation}
where $M$ is the scale parameter, the beta function is defined as $\beta_{\bar g} = M \frac{\partial \bar g }{ \partial M}$, and $\bar\gamma$ is the anomalous dimension. RG invariance in YMF theory means that, after proper scaling, the loop in a higher-order graph in perturbative expansion is identical to a loop in lower orders. This is a direct consequence of the CS equation, and it is indicative of the self-similar properties of gauge fields. These properties have important consequences for the dynamics of partons, as we will see below.

\begin{figure*}[t]
 \begin{tabular}{c@{\hspace{3.5em}}c}
\hspace{2cm} \includegraphics[width=0.30\textwidth]{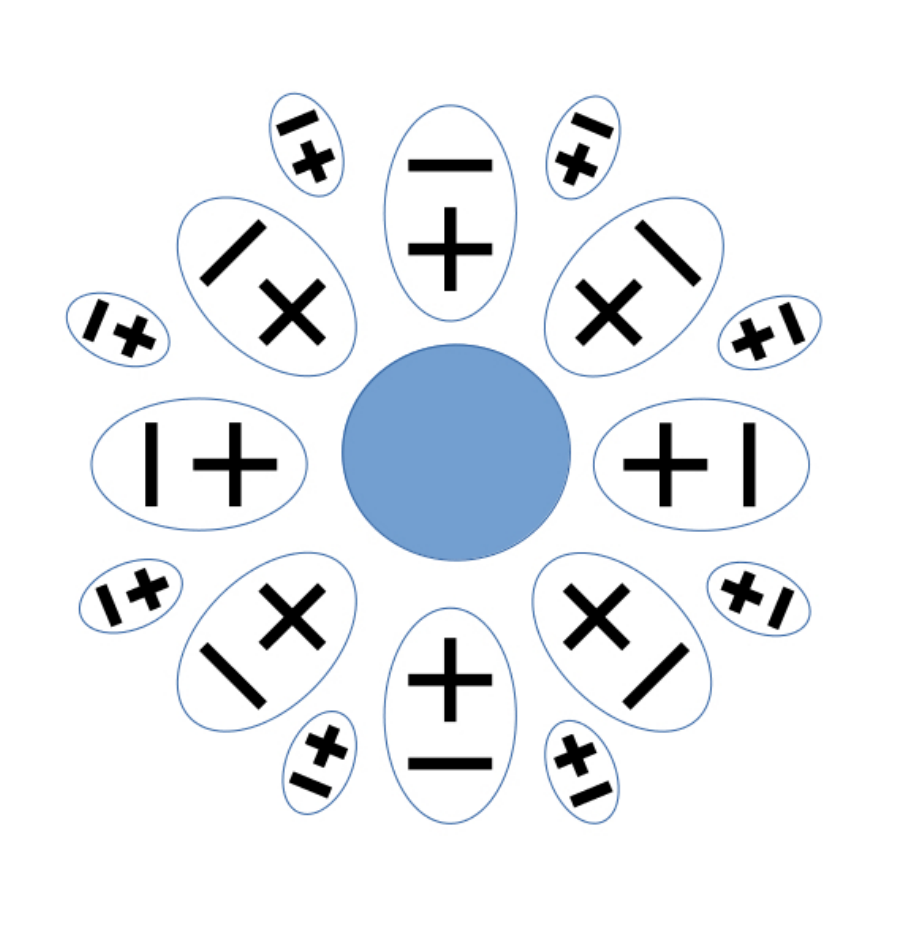} &
\includegraphics[width=0.40\textwidth]{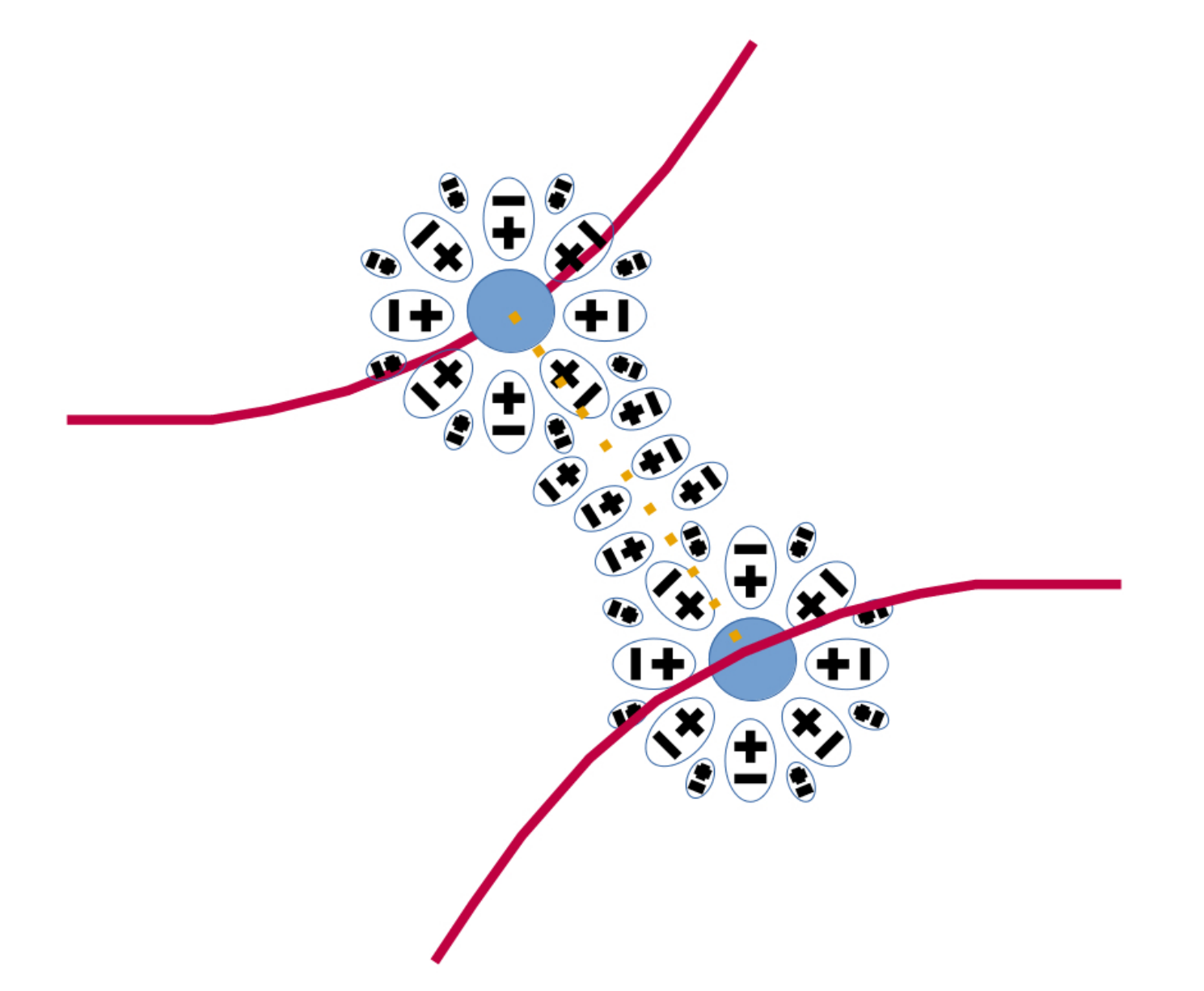}
\end{tabular}
$\qquad\qquad\qquad\qquad\qquad$ (a)         \hspace{6.2cm}   (b)
 \caption{\it Pictorial representation of the effective parton and their  interactions. (a)~The vacuum polarization represented as the internal structure of the effective parton. (b)~In the effective parton interaction the vacuum structure participates in a complex way.}
\label{fig:effective_parton}
\end{figure*}
The partonic dynamics can be described by a Dyson-Schwinger
expansion~\cite{Dyson:1949ha}, leading to an effective parton which
includes the self-energy interaction in the propagation of the
elementary parton. Then, RG invariance is responsible for a complex
structure of the effective parton which is depicted in
Fig.~\ref{fig:effective_parton}~(a). In this figure, the vacuum polarization is represented by the $+$ and $-$ signs surrounding the elementary parton, represented by the central circle. In this sense, we
say that the effective parton has an internal structure. The complexity of this structure can be evaluated by the number of Feynman graphs necessary to describe the self-interaction contributions even
in low-orders of calculation. The proper-vertex interaction is still more complex, as can be observed in Fig.~\ref{fig:effective_parton}~(b). The interaction is mediated by another parton (boson) which has its own self-energy contributions. The detailed description of all possible configurations
is a huge challenge to perturbative QCD. The present situation has led some authors to claim that the perturbative QCD approach will not be able to provide an accurate calculation of the running coupling constant at low energies, and that the renormalization procedure just exchanged the infinities of the vertex functions by an infinity number of parameters in the calculation of this constant.

By using the thermofractal ideas introduced in Sec.~\ref{subsec:thermofractals}, it has been derived in Refs.~\cite{Deppman:2016fxs,Deppman:2019yno} an effective description of YMF theory. We will summarize below the main results. Let us consider that the system with energy $E$, in which the parton with energy $\varepsilon_j$ is one among $N$ constituents, is itself a parton inside a larger system. Then, the power-law distribution of Eq.~(\ref{eq:PTFpm}) describes how the energy received by the initial parton flows to its internal d.o.f., i.e. to partons at higher perturbative orders. This suggests that at each vertex, this distribution plays the role of an effective coupling
\begin{equation}
\bar{g}= G \prod_{i=1}^{\tilde{N}} \left[1 + (q-1)\frac{\varepsilon_i}{k\tau}\right]^{-\frac{1}{q-1}} \,, \label{eq:geff}
\end{equation} 
where $\tilde{N}$ is the number of particles created or annihilated at each interaction, and $G$ is the overall strength of the interaction. Within this picture, the entropic index $q$ is related to the number of internal d.o.f. in the fractal structure.

\begin{figure*}[t]
\centering
 \begin{tabular}{c@{\hspace{3.5em}}c}
 \includegraphics[width=0.46\textwidth]{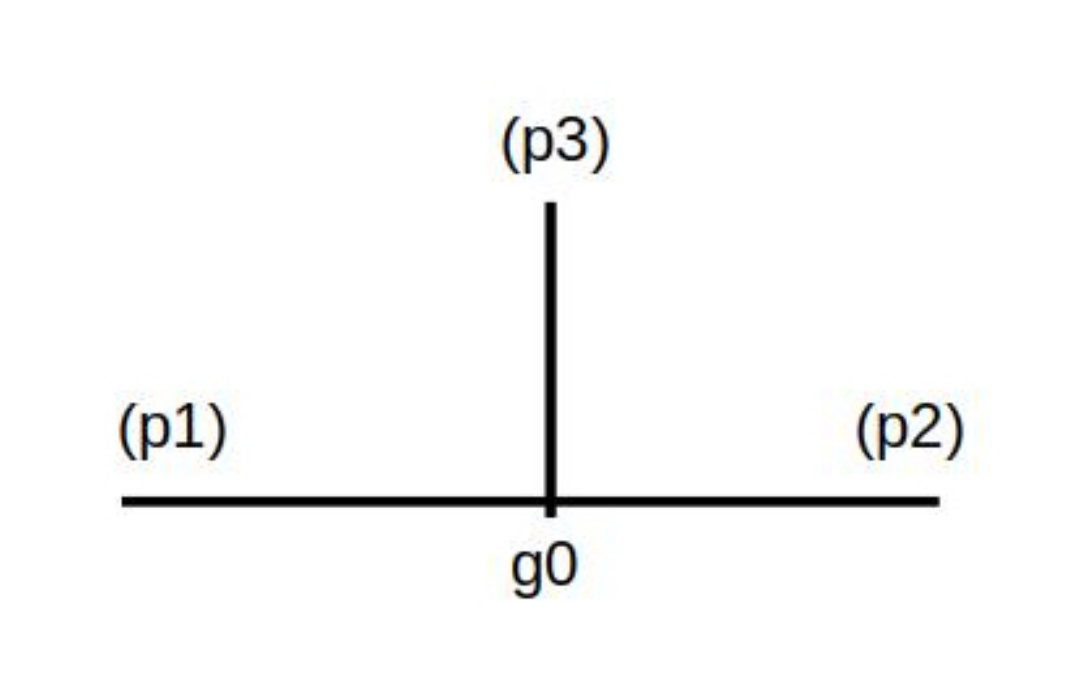} &
\includegraphics[width=0.42\textwidth]{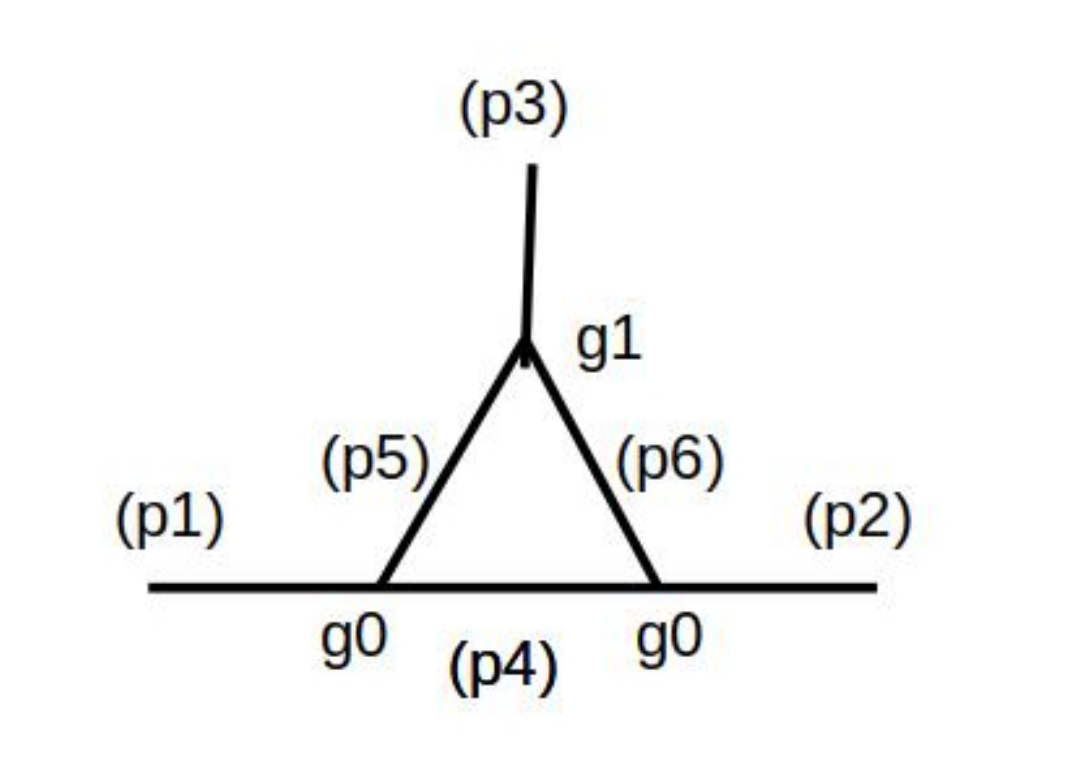}
\end{tabular}
 \caption{\it Vertex functions at scale $\lambda_0$ (left) and $\lambda$ (right).}
\label{fig:1loop}
\end{figure*}

The renormalized vertex functions together with the CS equation were used to derive the beta function of QCD, leading to the 1-loop result~\cite{Politzer:1974fr}
\begin{equation}
\beta_{\textrm{\scriptsize QCD}} = - \frac{1}{16\pi^2} \left[ \frac{11}{3}c_1 - \frac{4}{3} c_2 \right] \bar{g}^{3} \,, \label{eq:betaQCD}
\end{equation}
where $c_1 = N_c$ and $c_2 = N_f / 2$. The beta function can be derived as well by using the effective thermofractal description introduced above. To do this, one should consider a vertex at two different scales $\lambda_o$ and $\lambda$. As it is depicted in Fig.~\ref{fig:1loop}, the vertex function at scale $\lambda$ contains one additional loop, from which one can identify the effective coupling~$\bar g$. 
By using Eq.~(\ref{eq:geff}) with $\lambda = \lambda_o /\mu$, where $\mu$ is a scaling factor, the 1-loop beta function turns out to be
\begin{equation}
\beta_{\bar g} = \mu \frac{\partial \bar{g}}{\partial \mu}= - \frac{1}{16\pi^2} \frac{1}{q-1} g^{\tilde{N}+1} \,,
\end{equation}
with $\tilde{N} = 2$ in YMF theory.  Finally, from a comparison with the QCD result of Eq.~(\ref{eq:betaQCD}), one can relate the entropic index $q$ with the gauge field parameters, leading to~\cite{Deppman:2019yno,Deppman:2020gbu}

\begin{equation}
q = 1 + \frac{3}{11 N_c - 2 N_f} \,.
\end{equation}
This leads to $q \simeq 1.14$ when using $N_c = 3$ and $N_f = 6$, in excellent agreement with the experimental data analyses as we will see in the next section.

\section{Tsallis statistics: applications to high energy and hadron physics}
\label{sec:applications}

We will discuss in this section some of the recent applications of Tsallis statistics to QCD phenomenology, including HEP, hadron physics and BEC.

\subsection{Transverse momentum distribution in $pp$ collisions}
\label{subsec:pT}

\begin{figure*}[t]
\centering
 \begin{tabular}{c@{\hspace{2.5em}}c}
 \includegraphics[width=0.42\textwidth]{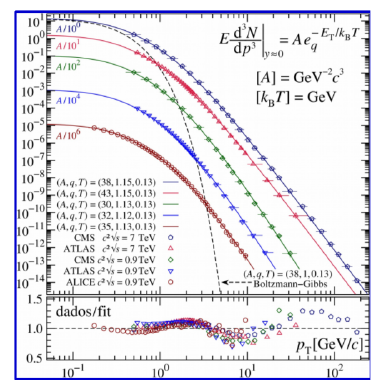}  &
 \includegraphics[width=0.50\textwidth]{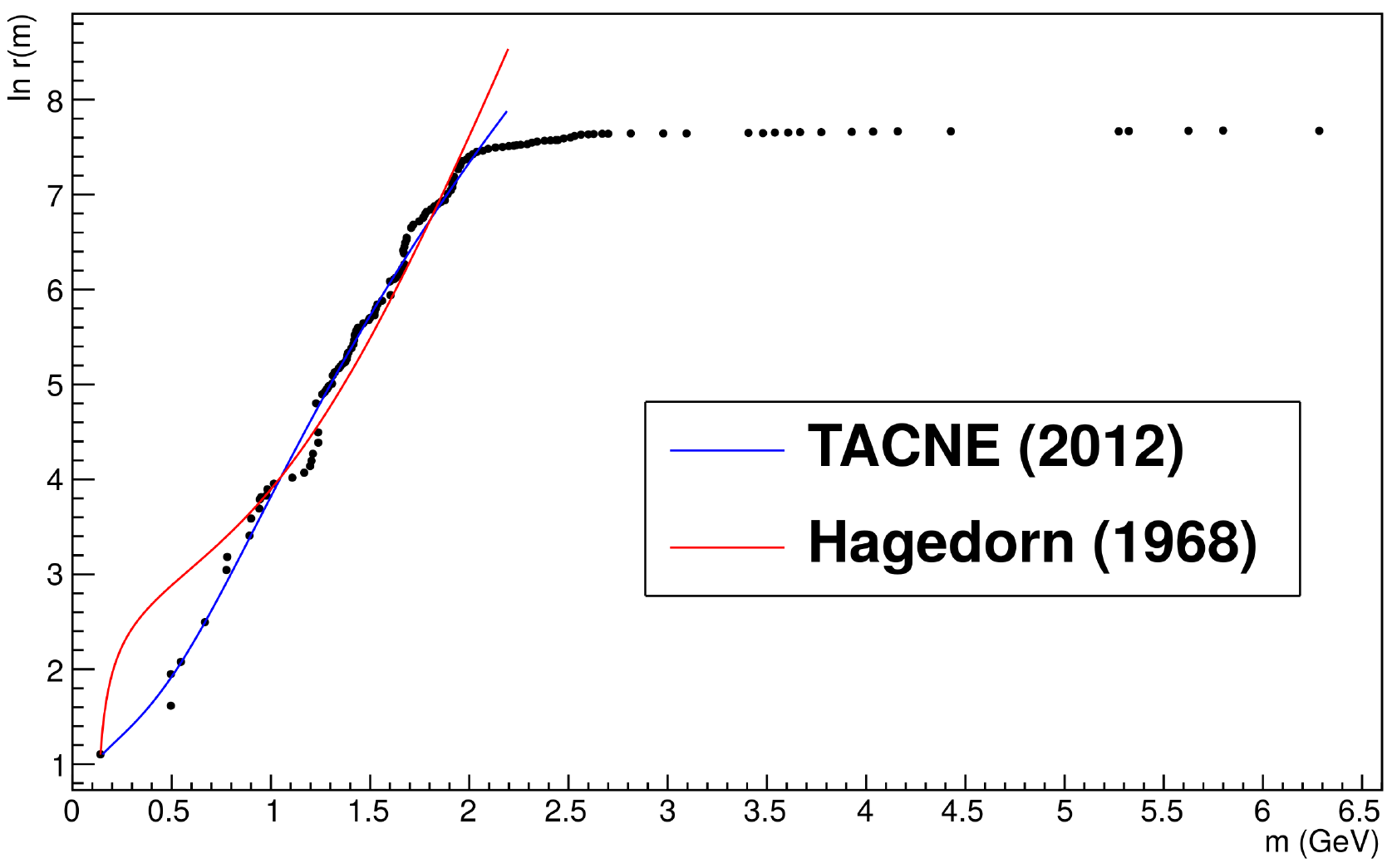} 
\end{tabular}
 \caption{\it Left panel: Fitting of the experimental data for $p_\perp$ distribution of the abundance of different hadron species in $pp$ collisions, by considering the NESCT distribution of Eq.~(\ref{eq:pt_dist_Tsallis}). Right panel: Cumulative hadron spectrum, as a function of the hadron mass. The dots stand for the PDG result~\cite{ParticleDataGroup:2020ssz}. We display also the result by using the NESCT (blue) and the result predicted by Hagedorn (red), cf. Ref.~\cite{Marques:2012px}.}
\label{fig:pp}
\end{figure*}

R. Hagedorn proposed a self-consistent thermodynamical approach to QCD formulated in terms of BG statistics, known as the HRG approach, allowing a description of the confined phase as a multi-component gas of non-interacting massive stable and point-like particles~\cite{Hagedorn:1965st,Hagedorn:1984hz}. When Hagedorn's theory was applied to $pp$ collisions, it predicted the transverse momentum distribution of the particle production of hadrons given by
\begin{equation}
\frac{d^2 {\mathcal N}}{dp_\perp dy}  = g V \frac{p_\perp m_\perp}{(2\pi)^2} e^{-\beta m_\perp}\,, \label{eq:pt_dist_BG}
\end{equation}
where $g$ is a constant, $\beta \equiv 1/(k_B T)$, $V$ is the volume of the system, $m_\perp = (p_\perp^2 + m^2)^{1/2}$, and $y$ is the rapidity. However, this exponential distribution turned out to be in disagreement with experimental data, as these behave instead as a power-law, cf. Fig.~\ref{fig:pp} (left). This was the motivation to consider the extension of Hagedorn's theory to non-extensive statistics, within the so-called non-extensive self-consistent thermodynamics (NESCT)~\cite{Deppman:2012us}. In this formalism, the $p_\perp$ distribution of particle species in $pp$ collision turns out to be
\begin{equation}
\frac{d^2 {\mathcal N}}{dp_\perp dy}  = g V \frac{p_\perp m_\perp}{(2\pi)^2} e_q^{(-)}\left( -\beta m_\perp \right) \,. \label{eq:pt_dist_Tsallis}
\end{equation}
This extended theory allows to reproduce the distribution of all the hadron species with high accuracy over $15$ orders of magnitude, leading to $q=1.14(1)$ and $T = 62(5) \, \textrm{MeV}$~\cite{Marques:2012px,Marques:2015mwa}. 
A second prediction of the NESCT is a power-law behavior for the hadron spectrum, with a density of hadron species given by~\cite{Deppman:2012us}
\begin{equation}
\rho(m)= \rho_o \cdot e_q^{(+)}\left( \beta m \right) \,. \label{eq:rhom}
\end{equation}
We display in Fig.~\ref{fig:pp} (right) the cumulative number of hadrons, defined as the number of hadronic states below some mass $m$, i.e. $N(m) \equiv \int_0^m d{\tilde m} \, \rho(\tilde{m})$. It is found that the distribution of Eq.~(\ref{eq:rhom}) leads to an excellent description of the hadron spectrum taken from the review by the Particle Data Group (PDG)~\cite{ParticleDataGroup:2020ssz}, as compared to the exponential distribution $\rho(m) = \rho_o \cdot e^{m/T_H}$ proposed by Hagedorn, specially for the lightest hadrons, cf. Ref.~\cite{Marques:2012px}.

\subsection{QCD thermodynamics}
\label{subsec:QCD_thermodynamics}

Tsallis statistics has been applied also to study the thermodynamics of QCD. The grand-canonical partition function for a non-extensive ideal quantum gas is given by~\cite{Megias:2014tha,Megias:2015fra}
\begin{equation}
\ln Z_q(V,T,\mu) =
 -\xi V\int \frac{d^3p}{{(2\pi)^3}} \sum_{r=\pm}\Theta(r x)\ln^{(-r)}_q\left(\frac{ e_q^{(r)}(x)-\xi}{ e_q^{(r)}(x)}\right) \,, \label{eq:lnZq}
\end{equation}
where $x = \beta(\varepsilon_p - \mu)$, the particle energy is $\varepsilon_p = \sqrt{p^2 + m^2}$, $\mu$ is the chemical potential, $\xi = \pm 1$ for bosons(fermions), and $\Theta(z)$ is the step function. The partition function for bosons is defined only for the case $\mu \le m$, therefore the term with $r = -$ in the integrand is applied only for fermions, and it only contributes if $\mu > m$. 

The thermodynamics of QCD in the confined phase has been widely studied within the HRG approach in which physical observables are described in terms of hadronic degrees of freedom~\cite{Hagedorn:1984hz}. These are usually taken as the conventional hadrons listed in the PDG~\cite{ParticleDataGroup:2020ssz}. In this approach, the partition function is given~by
\begin{equation}
\ln Z_q(V,T,\{\mu_{Q_{a}}\})=\sum_{i \in \textrm{\scriptsize hadrons}} \ln Z_q(V,T,\mu_{Q_{ai}})\,, \label{eq:logZ}
\end{equation}
where $\mu_{Q_{ai}} \equiv \mu_{a} Q_{ai}$ refers to the chemical potential of charge $Q_a \equiv \{ u, d, s\}$ for the $i$-th hadron, while $\mu_a$ is the chemical potential associated to charge~$Q_a$~\footnote{While we are considering the flavour basis $\{u,d,s\}$ of the $N_f = 3$ flavor sector of QCD, where $u$ refers to the number of constituent quarks minus antiquarks of type $u$ (and similarly for $d$ and $s$), we could work equivalently in the basis of conserved charges formed by the baryon number $B$, electric charge $Q$, and strangeness~$S$.}. From that, one can compute the thermodynamic quantities by using the standard thermodynamics relations. The thermal expectation value for the charge $Q_a$ is given by
\begin{equation}
\langle Q_a \rangle = \frac{1}{\beta} \frac{\partial}{\partial \mu_{a}} \ln Z_q \bigg|_{\beta}  = Q_a \langle N_q \rangle \,, \label{eq:Qa}
\end{equation}
where $\langle N_q \rangle$ is the average number of particles. By using that the baryon number for (anti)quarks is $B_{\textrm{\scriptsize quarks}} = 1/3$ and $B_{\textrm{\scriptsize antiquarks}} = -1/3$, the baryon density turns out to be
\begin{equation}
\rho_B = \frac{\langle B\rangle}{V} = \frac{1}{3V} \left( \langle N_{\textrm{\scriptsize quarks}}\rangle - \langle N_{\textrm{\scriptsize antiquarks}}\rangle \right) \,. \label{eq:rhoB}
\end{equation}
The thermodynamical relations for the pressure $P$, energy density $\varepsilon$, and entropy $S$, involve derivatives of $\ln Z_q$ with respect to $V$, $\mu_B$ and $\beta$~\cite{Megias:2015fra,Andrade:2019dgy}.
\begin{figure*}[t]
\centering
 \begin{tabular}{c@{\hspace{2.5em}}c}
 \includegraphics[width=0.46\textwidth]{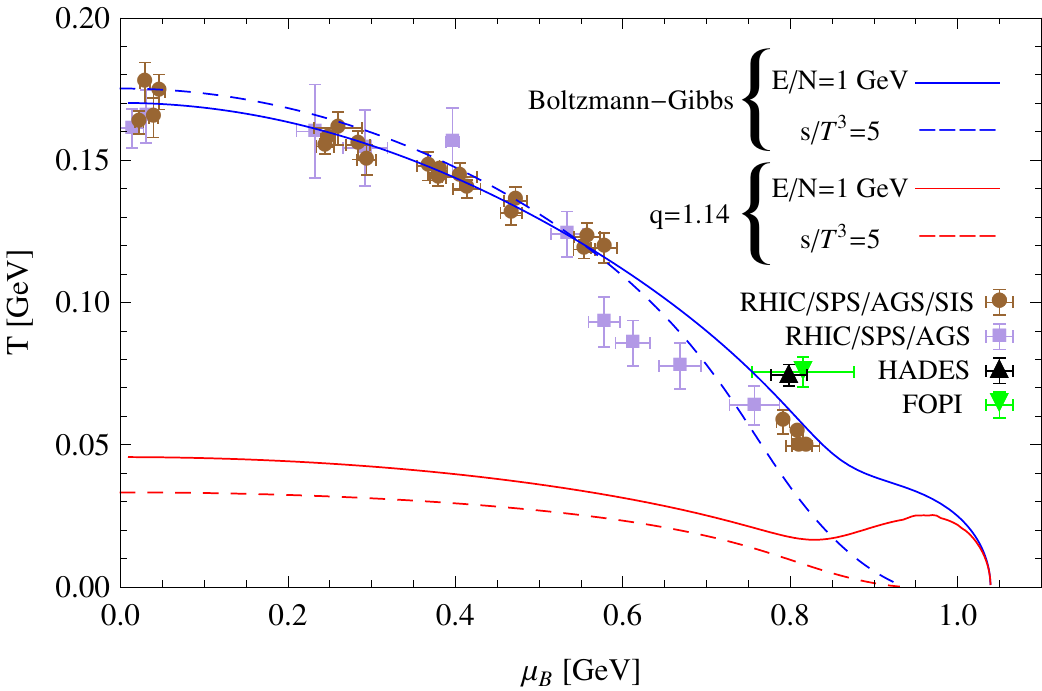} &
\includegraphics[width=0.45\textwidth]{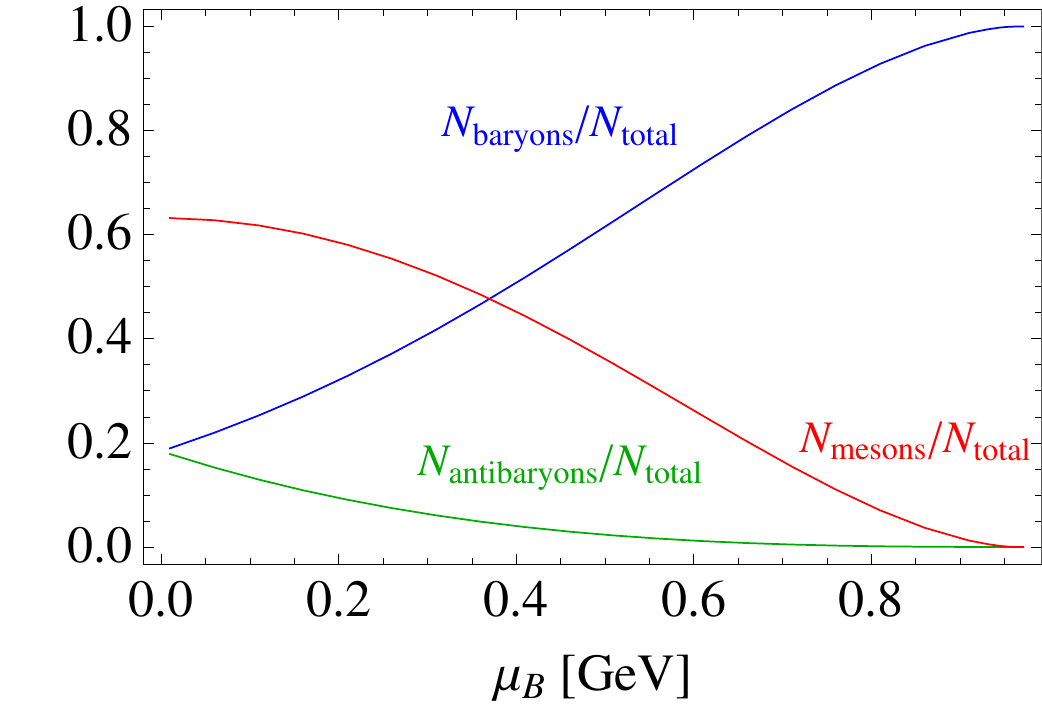}
\end{tabular}
 \caption{\it Left panel: Chemical freeze-out line $T = T(\mu_B)$. We plot the result by using BG statistics, and Tsallis statistics. Right panel: Number of (anti)baryons/mesons inside a volume $V_{\textrm{\scriptsize proton}}$, as a function of $\mu_B$. We have considered in both panels $q = 1.14$.}
\label{fig:TmuS}
\end{figure*}
Using the arguments of Refs.~\cite{Cleymans:1999st,Tawfik:2005qn}, the chemical~freeze-out~line~$T = T(\mu_B)$ can been determined by the conditions $\langle E\rangle/\langle N \rangle \simeq 1 \, \textrm{GeV}$ or $s/T^3  \simeq 5$. The results, displayed in Fig.~\ref{fig:TmuS} (left), show an inflection for $\mu_B \simeq m_{\textrm{\scriptsize proton}}$ related to a sharp increase of the baryon density in this regime. The region below the freeze-out line refers to the confined regime.

One important aspect to study is the limits of temperature and chemical potential within which the proton can exist as a confined system. To address this point, we can consider the MIT-bag model criterion, i.e. the proton exists only if the total energy inside a volume $V_{\textrm{\scriptsize proton}}$ is smaller or equal to the proton mass, $\varepsilon \cdot V_{\textrm{\scriptsize proton}} \le m_{\textrm{\scriptsize proton}}$~\cite{Andrade:2019dgy}. Fig.~\ref{fig:TmuS} (right) shows the number of baryons, antibaryons and mesons normalized to the total number of hadrons, along the line $\varepsilon \cdot V_{\textrm{\scriptsize proton}} = m_{\textrm{\scriptsize proton}}$. According to this figure, the proton exists close to $\mu_B \simeq m_{\textrm{\scriptsize proton}}$, and at this value of chemical potential, the proton is completely baryonic in content.

\begin{figure*}[t]
\centering
 \begin{tabular}{c@{\hspace{3.5em}}c}
 \includegraphics[width=0.46\textwidth]{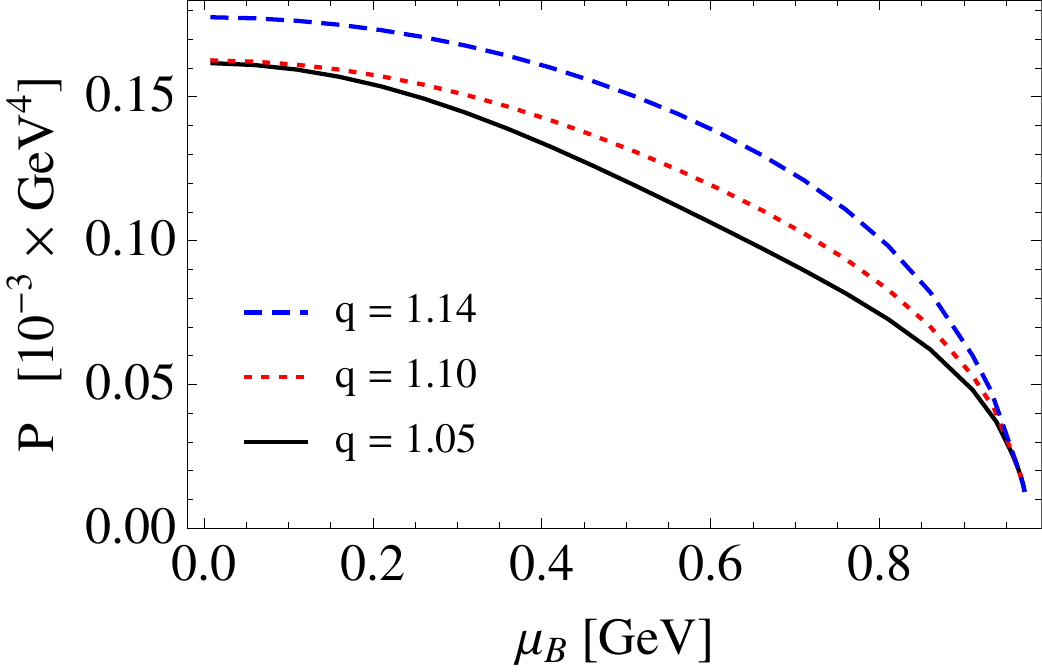} &
\includegraphics[width=0.46\textwidth]{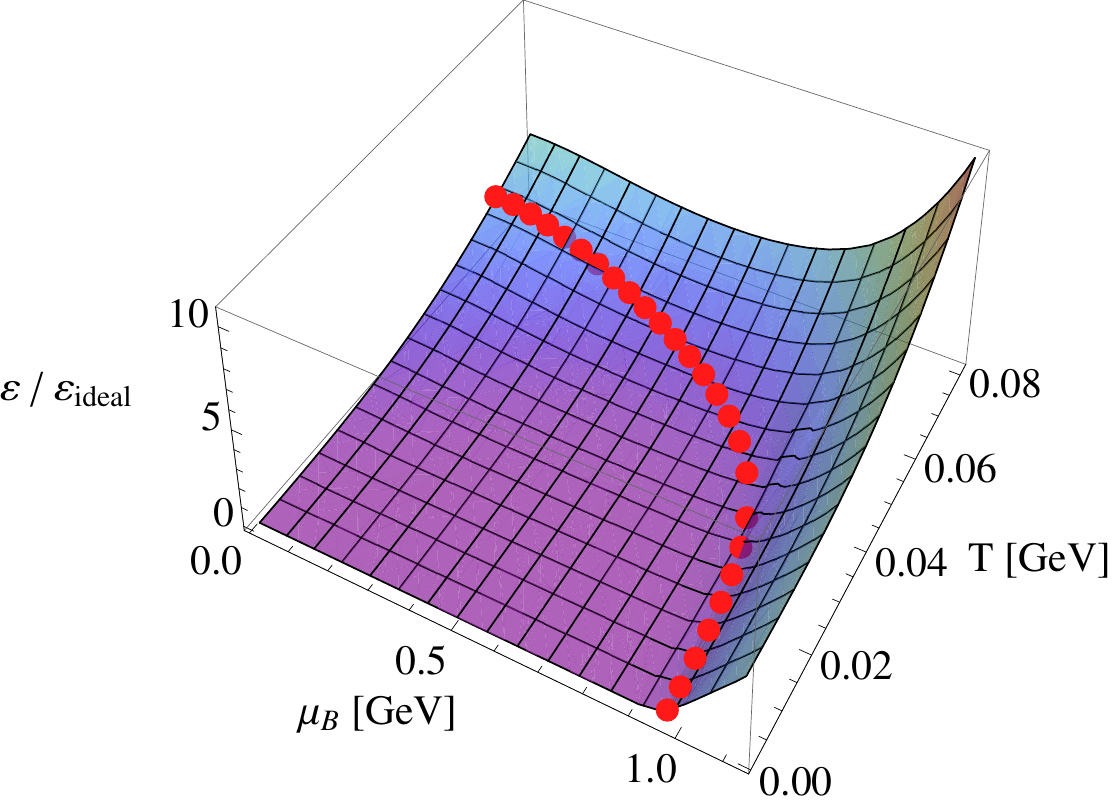}
\end{tabular}
 \caption{\it Left panel:  Pressure as a function of the baryonic chemical potential for different values of $q$. The temperature $T$ is chosen to keep the total energy $\varepsilon \cdot V_{\textrm{\scriptsize proton}}$ fixed to the value $ m_{\textrm{\scriptsize proton}}$. Right panel: Energy density (normalized to the massless ideal gas limit) in the $(\mu_B,T)$ plane. The red dots corresponds to the region in which $\varepsilon \cdot V_{\textrm{\scriptsize proton}} = m_{\textrm{\scriptsize proton}}$. We have considered $q = 1.14$.}
\label{fig:EP}
\end{figure*}

In order to evaluate the effects of non-extensivity in the thermodynamic quantities, we present in Fig.~\ref{fig:EP} (left) a plot of the pressure as a function of $\mu_B$ for different values of the entropic index~$q$. As it was discussed in Refs.~\cite{Megias:2015fra,Andrade:2019dgy}, the equation of state $P = P(\varepsilon)$ becomes harder for $q > 1$ as compared to BG statistics. This has important implications for neutron stars, in particular, the non-extensive effects turn out to be enough to produce stars with higher maximum masses~\cite{Menezes:2014wqa}.
Finally, we display in Fig.~\ref{fig:EP} (right) the energy density in the $(\mu_B,T)$ plane. The curve for which $T$ and $\mu_B$ results in total energy equal to the proton mass is indicated by red points. The system seems to behave close to the conformal limit in this regime, so that the trace anomaly is vanishing, i.e. $\varepsilon - 3P \simeq 0$. Using that $\varepsilon \cdot V_{\textrm{\scriptsize proton}} = m_{\textrm{\scriptsize proton}}$ together with $P \simeq \varepsilon/3$, one finds
\begin{equation}
P = \frac{m_{\textrm{\scriptsize proton}}}{3 V_{\textrm{\scriptsize proton}}} = (0.135 \, \textrm{GeV})^4 \,. \label{eq:bag_constant}
\end{equation}
This value, which is interpreted as the bag constant of the model, is consistent with the vacuum energy density obtained from QCD calculations: $\varepsilon = (0.161 \, \textrm{GeV})^4$~\cite{Schaden:1998ty}. Common values in the literature of the bag constant lie in the range $(0.145 \,\textrm{GeV})^4 - (0.250\, \textrm{GeV})^4$~\cite{Cardoso:2017pbu,Rischke:1987xe}, so that the result of Eq.~(\ref{eq:bag_constant}) is in good agreement with this range.

\subsection{Bose-Einstein condensation and Tsallis statistics}
\label{subsec:BE_condensation}

The possible formation of a BEC in high energy collisions and hadronic systems has been widely studied in the literature. In these studies the critical temperature of the phase transition from the confined to the deconfined quark regimes are associated to the formation of a condensate, see e.g. Refs.~\cite{Kharzeev:2006zm,Bautista:2019mts,Deb:2021gzg}. While the BEC has been exhaustively studied under the light of BG statistics, the same does not hold for Tsallis statistics. We will study below the BEC phenomenon in non-extensive statistics (qBEC). We will adopt the relativistic description, which can be straightforwardly restricted to the non-relativistic case, as it will be commented on below. 

\begin{figure*}[t]
\centering
 \begin{tabular}{c@{\hspace{3.5em}}c}
 \includegraphics[width=0.39\textwidth]{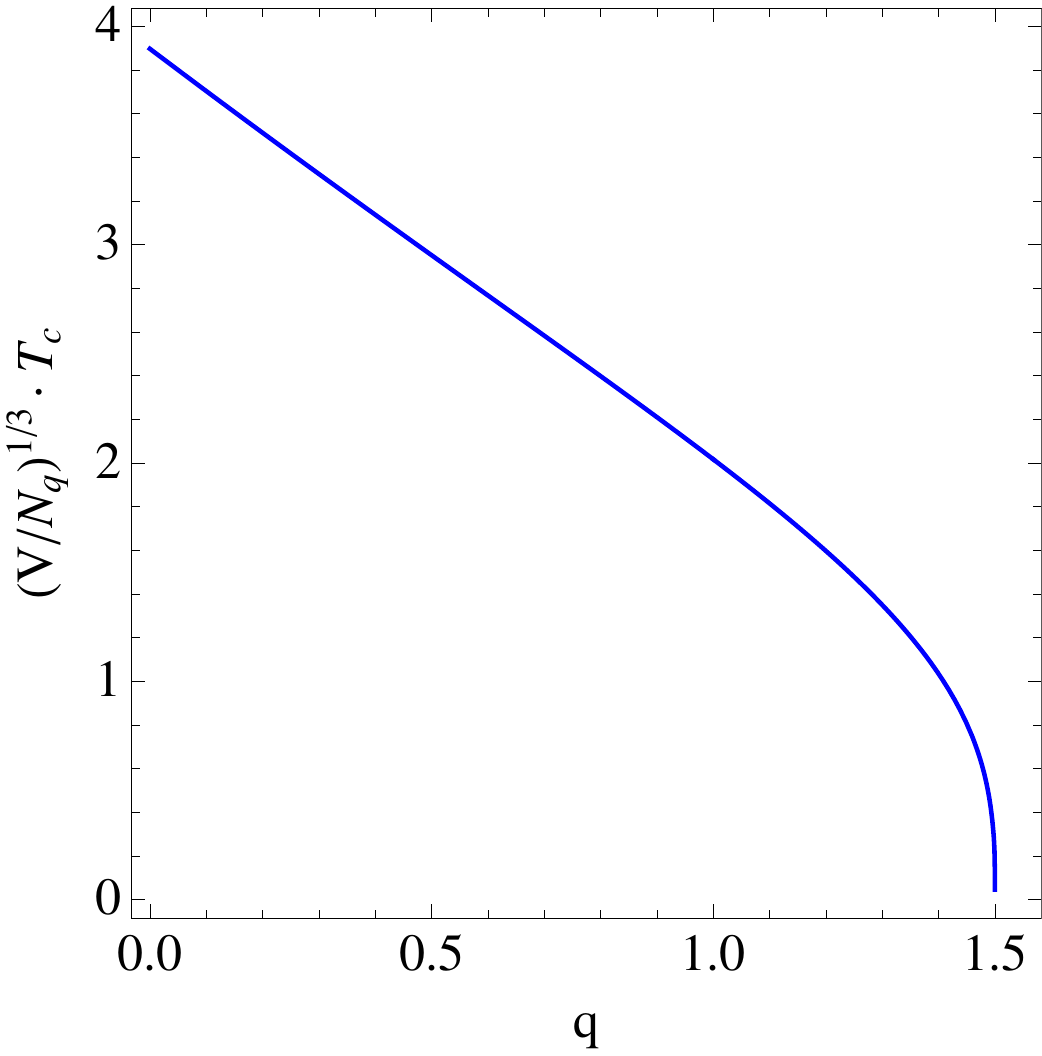} &
\includegraphics[width=0.42\textwidth]{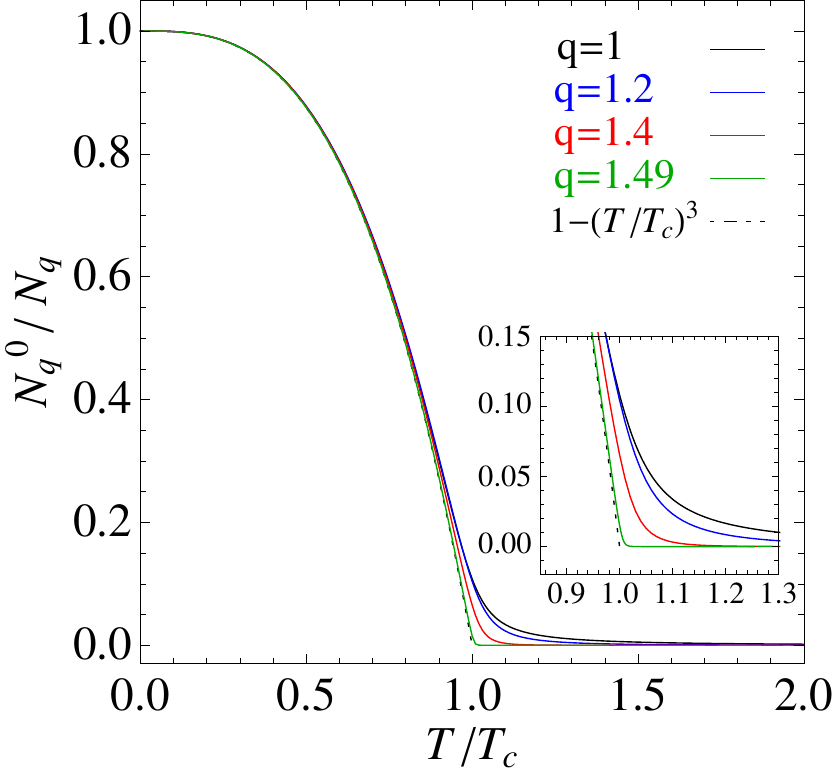}
\end{tabular}
 \caption{\it Left panel: Critical temperature $\left( \times (V/N_q)^{1/3} \right)$ as a function of the entropic index $q$. Right panel: BEC ratio as a function of $T/T_c$. This plot is for fixed value of the number of particles, $N_q = 100$, and different values of~$q$.}
\label{fig:Tc}
\end{figure*}

\begin{figure*}[t]
\centering
 \begin{tabular}{c@{\hspace{3.5em}}c}
 \includegraphics[width=0.40\textwidth]{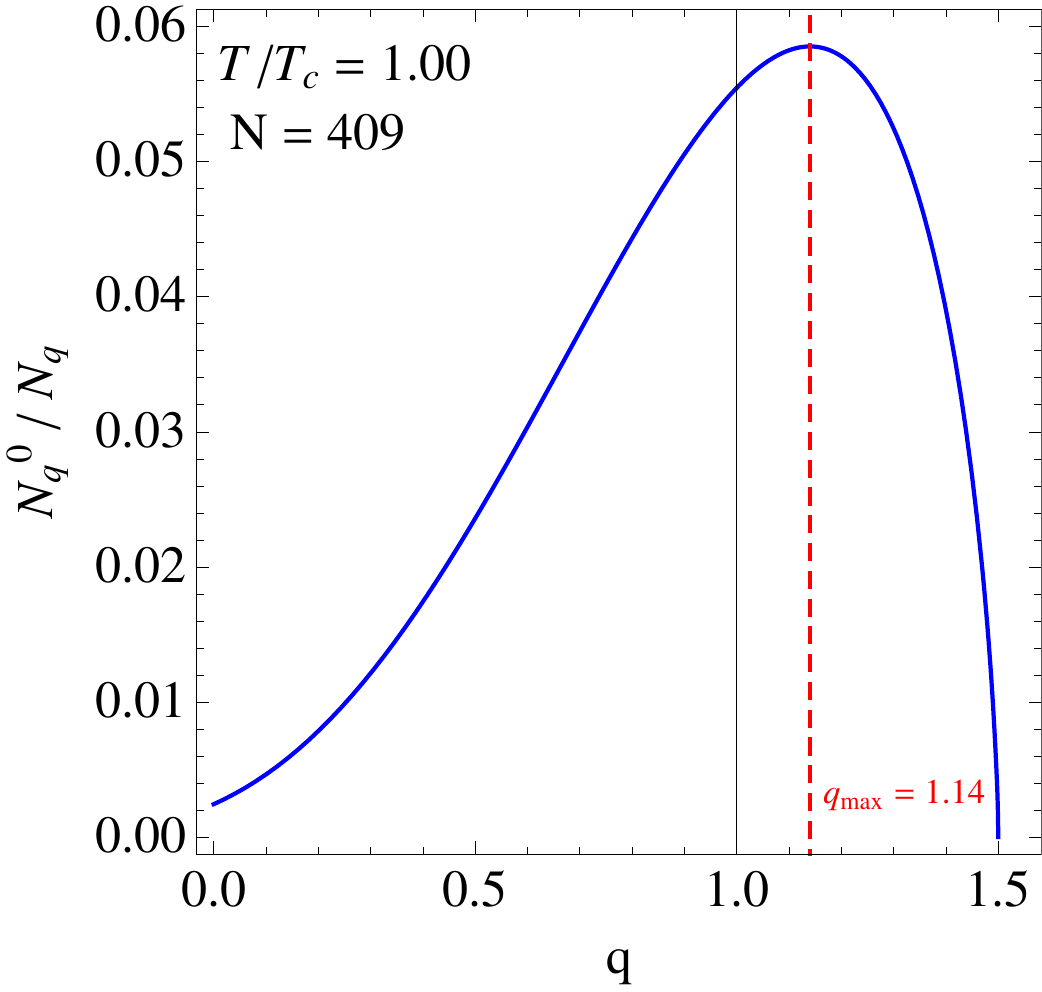} &
\includegraphics[width=0.42\textwidth]{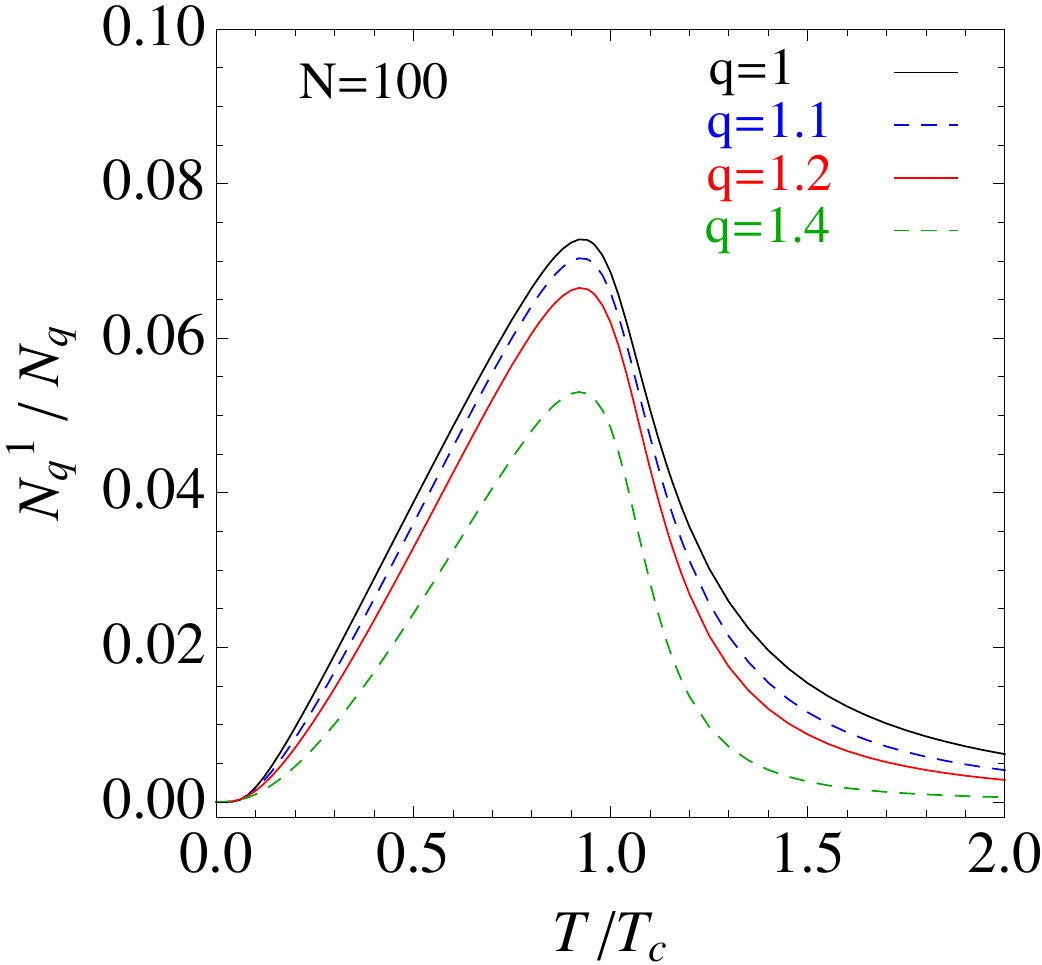}
\end{tabular}
 \caption{\it Left panel: Fraction of particles in the condensate at $T = T_c$ as a function of $q$. The maximum value of $N_q^0/N_q$ is obtained at $q_\textrm{\scriptsize max} = 1.14$ for $N_q = 409$. Right panel: Fraction of particles in the first excited state for $N_q = 100$, and different values of $q$.}
\label{fig:N0N1}
\end{figure*}

By using the grand-canonical partition function of Eq.~(\ref{eq:lnZq}), and the thermodynamical relation of Eq.~(\ref{eq:Qa}), the total number of particles of a relativistic non-extensive bosonic system is 
\begin{equation}
\hspace{-2.2cm} N_q \equiv N_q^0 + N_q^{\varepsilon} = \frac{1}{\left( e_q^{(+)}[\beta (\varepsilon_c-\mu)]-1 \right)^{q} } + \frac{V}{2 \pi^2 }  \int_0^\infty d\varepsilon \, \varepsilon^2 \frac{1}{\left( e_q^{(+)}[\beta (\varepsilon-\mu)]-1 \right)^{q} }  \,, \label{eq:Nq}
\end{equation}
where $N_q^0$ and $N_q^{\varepsilon}$ are the numbers of particles in the ground-state and excited states, respectively. If we consider $\mu \to 0$, the singularity in the occupation number corresponds to the ground-state, $\varepsilon_c = 0$. The BEC happens below some critical temperature, $T_c$, and it is signalled by a non-negligible value of $N_q^0$. When $\mu \to 0$, the maximum number of particles in the excited states is reached at the critical temperature, and is
\begin{equation}
\hspace{-2.2cm} N_{q,\textrm{\scriptsize max}}^{\varepsilon}(T_c) = \frac{V T_c^3}{\pi^2} \zeta_q(0) \,, \quad \textrm{where} \quad  \zeta_q(0) = \frac{1}{2} \int_{0}^\infty dx \, x^{2} \frac{1}{\left( e_q^{(+)}(x)  - 1 \right)^{q} }\,. \label{eq:zeta_q}
\end{equation}
Below the critical temperature, however, the number of particles allowed in the excited states becomes smaller than the maximum number of particles at the critical temperature, i.e. $N_q^\varepsilon(T) \le N_{q,\textrm{\scriptsize max}}^{\varepsilon}(T_c)$, so the excess of particles must be at the ground-state. If one considers that the total number of particles is $N_q = N_{q,\textrm{\scriptsize max}}^\varepsilon(T_c)$, which means $N_q^0(T_c) = 0$, then the critical temperature turns out to be~\cite{Megias:2021jji}
\begin{equation}
T_c = \left( \frac{N_q}{V} \right)^{1/3} \frac{\pi}{\left[ \pi \zeta_q(0) \right]^{1/3}} \,.
\end{equation}
We show in Fig.~\ref{fig:Tc} (left) the behavior of $(V/N_q)^{1/3} T_c$ with the entropic index~$q$. Notice that the curve is independent of the values of $V$ and $N_q$. The value of $T_c$ decreases with~$q$ up to the vicinity of the critical value $q_c = 3/2$, which represents the maximum value of $q$ for the formation of the BEC in non-extensive systems.
Below the critical temperature, the condensate ratio (fraction of particles in the ground state) is given by
\begin{equation}
\frac{N_q^0}{N_q} \simeq 1 - \left( \frac{T}{T_c} \right)^3 \,, \qquad (T \le T_c) \,.
\end{equation}
For a non-relativistic gas, we would have a similar equation with power $3/2$ in the temperature, instead of $3$. We display in Fig.~\ref{fig:Tc} (right) the results of $N_q^0/N_q$ as a function of $T/T_c$. The results of Fig.~\ref{fig:Tc} evidence an interesting behaviour of the qBEC that cannot be observed in BG statistics. We observe in the left panel the resistance of the system to form the condensate as $q$ increases, which is manifested in the lower critical temperatures. On the other hand, the right panel shows that the phase transition to the condensate is sharper for systems with larger values of $q$. We display in Fig.~\ref{fig:N0N1} (left) the dependence with the entropic index $q$ of the condensate ratio at the critical temperature. We observe a peak in the curve at a position $q_{\textrm{\scriptsize max}}$, which depends on the number of particles in the system. The numerical analysis shows that $q_{\textrm{\scriptsize max}} = 1.14$ is obtained for $N_q = 409$.

To investigate these features in more detail, we can study the fraction of particles in the first excited state. While the system of Eq.~(\ref{eq:Nq}) is supposed to have a continuum of states, one can obtain a discretization of the energy levels when considering it inside a large cubical box of length $L$. Then, the energy levels of the relativistic massless particles are  
\begin{equation}
E_{n_x, n_y, n_z} = \frac{\pi}{L}  \sqrt{n_x^2 + n_y^2 + n_z^2}  \,, \qquad n_x, n_y, n_z \ge 1  \,. \label{eq:Ebox}
\end{equation}
The results of $N_q^1/N_q$ obtained with this method are plotted in Fig.~\ref{fig:N0N1} (right). We see that there is a peak in this ratio close to the phase transition. The reduction of the number of particles in the first excited state for higher values of $q$ is associated with the fact that a larger fraction of the particles is in the ground state, leading to a sharper phase transition. This confirms the conclusions obtained above.

\begin{figure*}[t]
\centering
 \begin{tabular}{c@{\hspace{1.5em}}c@{\hspace{1.5em}}c}
 \includegraphics[width=0.30\textwidth]{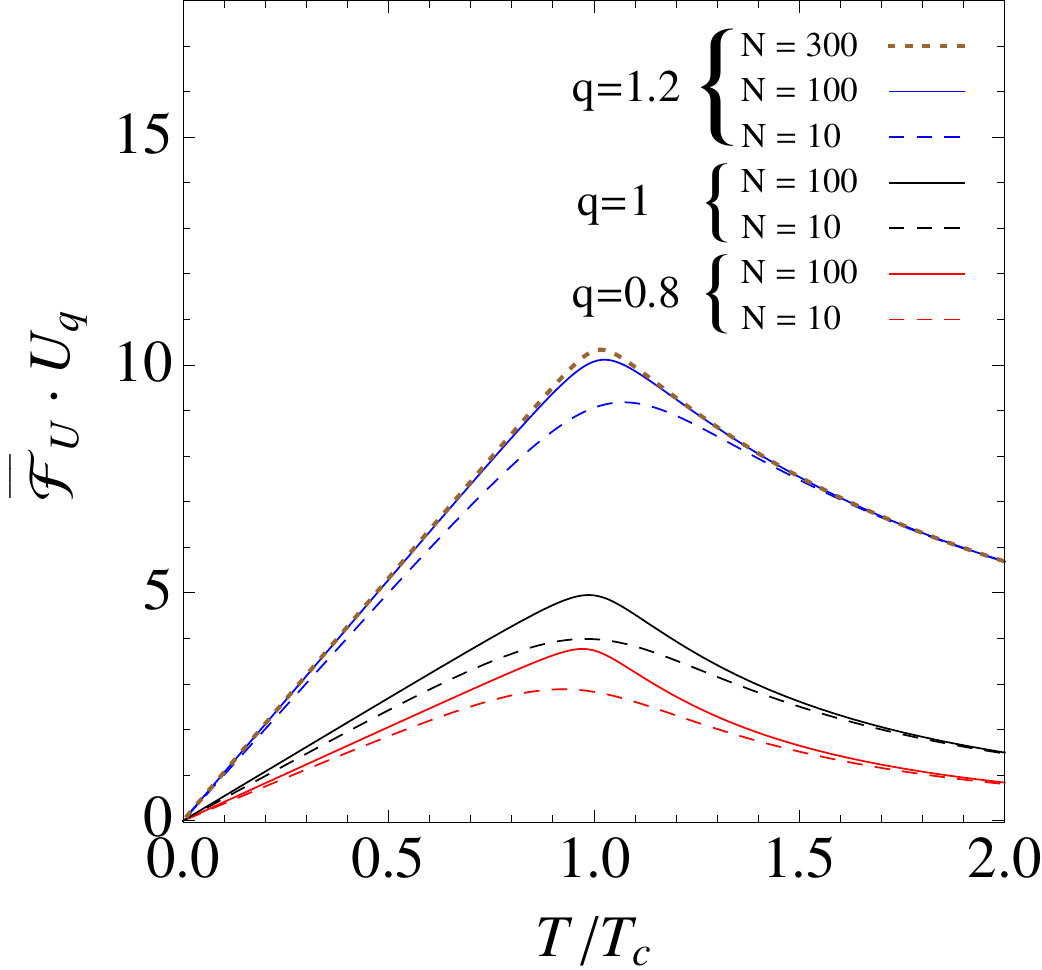} &
 \includegraphics[width=0.29\textwidth]{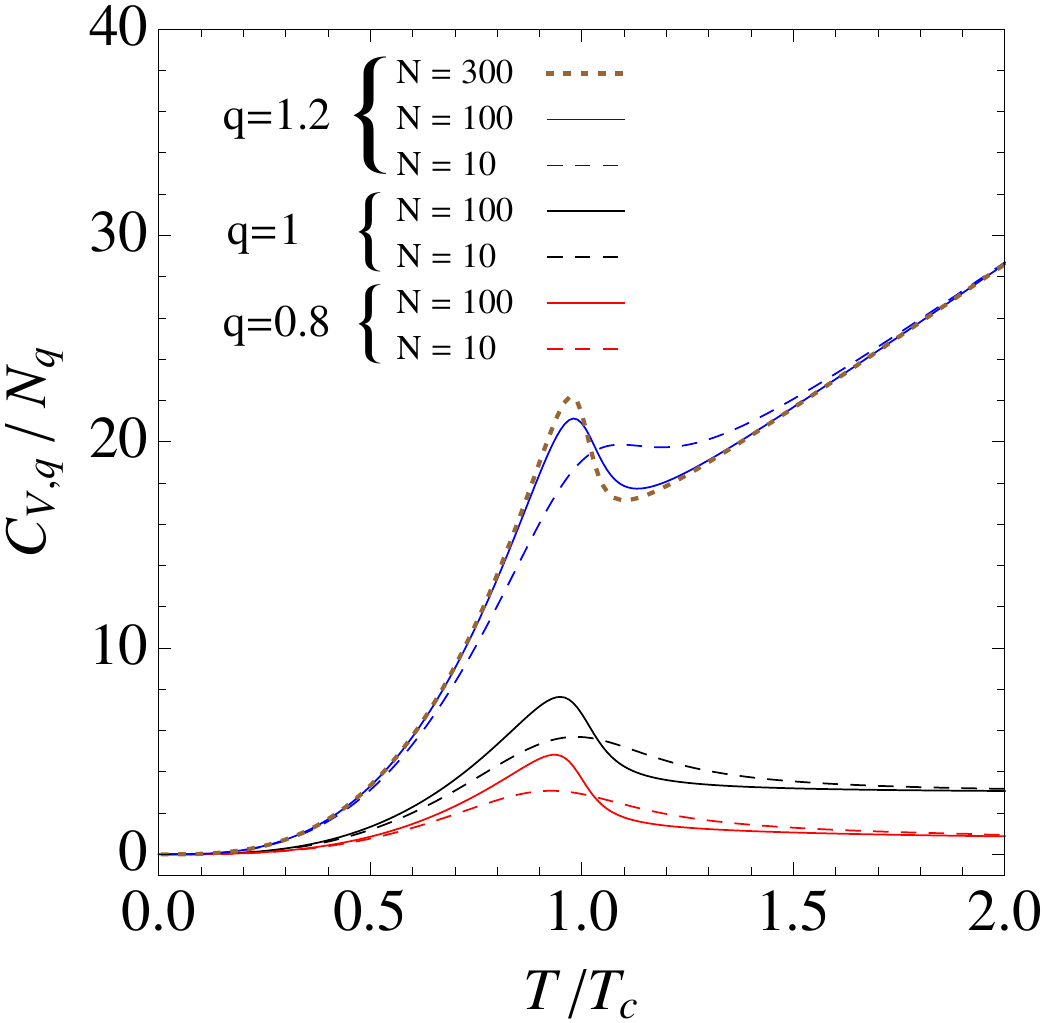} &
 \includegraphics[width=0.31\textwidth]{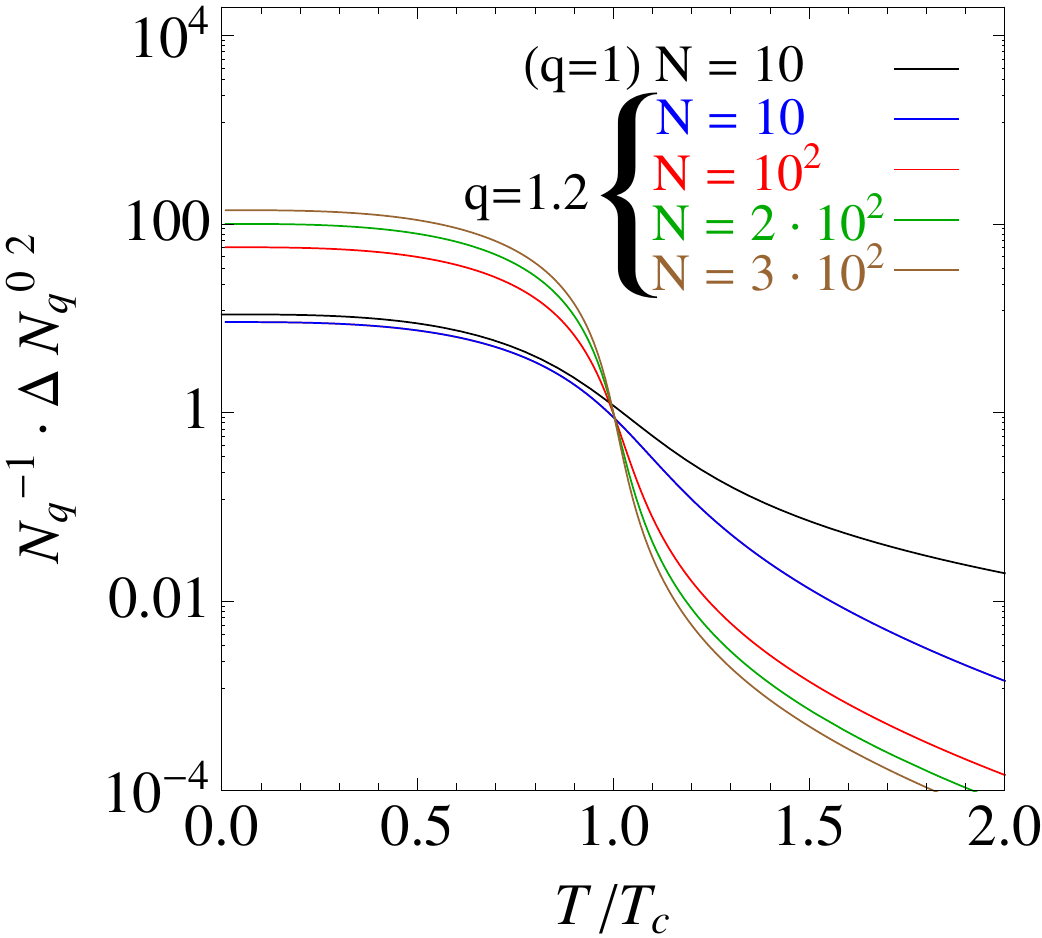}
\end{tabular}
 \caption{\it Total energy, normalized by the factor $\overline{\mathcal F}_U = \frac{1}{N_q} \left( \frac{V}{N_q} \right)^{1/3} \left( \frac{T_c}{T} \right)^3$ (left panel), specific heat at constant volume $(\times N_q^{-1})$ (middle panel), and variance of the condensate population $(\times N_q^{-1})$ (right panel). We display the results for different values of $q$ and~$N_q$.}
\label{fig:UqCVDeltaN}
\end{figure*}

We have studied other thermodynamical quantities, in particular, the total energy $U_q \equiv \langle E \rangle = \varepsilon \cdot V$, the specific heat
\begin{equation}
C_{V, q} \equiv \frac{\partial U_q}{\partial T}  \Big|_{N_q,V} \,,
\end{equation}
and the variance of the condensate population
\begin{equation}
\Delta N_q^{0 \, 2} \equiv  \left\langle  \left( N_q^{0} - \langle N_q^{0}\rangle \right)^2 \right\rangle = \beta^{-1} \frac{\partial}{\partial \mu} N_q^0  \,.
\end{equation}
The dependence of these quantities with $T/T_c$ is displayed in Fig.~\ref{fig:UqCVDeltaN}.
We see in the middle panel of this figure that $C_{V,q} \propto T^3$ up to the critical temperature, while there is a change of regime for $T \gtrsim T_c$. We have checked that, when considering the thermodynamic limit, $N_q \to \infty$,  this smooth change becomes a discontinuity with $\Delta C_{V,q}(T_c) = C_{V,q}(T\gtrsim T_c) - C_{V,q}(T\lesssim T_c)  < 0$, as it was observed in Ref.~\cite{Chen:2002hx}. Finally, let us mention that the variance tends to decrease with the value of $q$, as it is shown in Fig.~\ref{fig:UqCVDeltaN} (right). This is an interesting quantity since it can be measured experimentally~\cite{Huang:2021vxw}. 

The physics of the non-relativistic qBEC can be studied similarly. The range of values for the entropic index $q$ where the qBEC can be obtained in this case is $0 < q < 3$. The differences between the relativistic and non-relativistic cases are due to the topology of the phase space.

\section{Conclusions}
\label{sec:conclusions}

In this work, we have reviewed recent applications of non-extensive statistics in the form of Tsallis statistics to HEP and hadron physics. These include the physics of high energy $pp$ collisions~\cite{Cleymans:2011in,Wong:2015mba,Marques:2015mwa}, hadron models~\cite{Cardoso:2017pbu,Andrade:2019dgy}, hadron mass spectrum~\cite{Marques:2012px}, QCD thermodynamics and neutron stars~\cite{Menezes:2014wqa}, and BEC~\cite{Chen:2002hx,Megias:2021jji}. Other applications not analyzed in this manuscript include heavy-ion collisions~\cite{DeppmanNunes:inprep}, hadron structure~\cite{DeppmanTeixeira:inprep}, lattice QCD~\cite{Deppman:2012qt}, and many other aspects of non-extensive statistical mechanics~\cite{Deppman:2012us,Megias:2015fra}. We have also investigated the structure of a thermodynamical system presenting fractal properties, showing that it naturally leads to Tsallis non-extensive statistics. Based on the self-similar properties of thermofractals, we have explained how a field theoretical approach for thermofractals can account for the dynamics of effective partons, and correctly reproduces the beta function of QCD, leading to a value of the entropic index $q \simeq 1.14$ which turns out to be in excellent agreement with phenomenological analyses~\cite{Deppman:2019yno,Deppman:2020gbu,Deppman:2020jzl}.

There are still many open questions. Regarding the description of HEP data by power-law distributions, the main problem is to verify to what extent the idea of fractal structure can describe experimental data, including analyses of the fractal dimension that can be accessed through intermittency analysis~\cite{Gupta:2015qja,Deppman:2020jzl}. These analyses can be eventually extended also for heavy-ion collisions~\cite{Hegyi:1992dj}.

Beyond the phenomenological success of Tsallis statistics and the thermofractal description, let us remark that self-similarity in gauge fields leads to interesting properties, as e.g. self-consistency and fractal structure, recursive calculations at any order, non-extensive statistics, reconciliation of Hagedorn's theory with QCD, and excellent agreement with experimental data for $pp$ and heavy-ion collisions. The study of all these features deserves further investigation.

\ack 

The work of E.M. is supported by the project PID2020-114767GB-I00
financed by MCIN/AEI/10.13039/501100011033, by the FEDER/Junta de
Andaluc\'{\i}a-Consejer\'{\i}a de Econom\'{\i}a y Conocimiento
2014-2020 Operational Program under Grant A-FQM178-UGR18, by Junta de
Andaluc\'{\i}a under Grant FQM-225, and by the Consejer\'{\i}a de
Conocimiento, Investigaci\'on y Universidad of the Junta de
Andaluc\'{\i}a and European Regional Development Fund (ERDF) under
Grant SOMM17/6105/UGR. The research of E.M. is also supported by the
Ram\'on y Cajal Program of the Spanish MCIN under Grant
RYC-2016-20678. A.D., D.P.M. and T.N.dS. are supported by the Project
INCT-FNA (Instituto Nacional de Ci\^encia e Tecnologia - F\'{\i}sica
Nuclear Aplicada) Proc.  No. 464898/2014-5. A.D. is partially
supported by the Conselho Nacional de Desenvolvimento Cient\'{\i}fico
e Tecnol\'ogico (CNPq-Brazil), grant 304244/2018-0, by Project INCT-
FNA Proc. No. 464 898/2014-5. A.G. is supported by CNPq, Brazil grant
PQ 306920/2018-2. AD and AG are supported by FAPESP, Brazil grant
2016/17612-7. V.S.T. is supported by FAEPEX (grant 3258/19), FAPESP
(grant 2019/010889-1) and CNPq (grant 306615/2018-5).

\section*{References}

\end{document}